\documentclass{jac}

\usepackage{graphicx}
\usepackage{amsfonts}
\usepackage{amsmath}
\usepackage{latexsym}
\usepackage{amssymb}
\usepackage[mathcal]{euscript}
\usepackage[table]{xcolor}
\usepackage{hyperref}

\newcommand\VL[1]{} 	
\newcommand\VS[1]{#1} 	

\newcommand{\Z}{\mathbb{Z}}

\newcommand{\pa}[1]{\left(#1\right)}

\newcommand{\eg}{\emph{e.g.\ }{}}
\newcommand{\ie}{\emph{i.e.\ }{}}

\newcommand{\ket}[1]{\mathinner{|{#1}\rangle}}
\newcommand{\Ket}[1]{\left|#1\right>}

\newcommand{\Phase}{R($\frac{\pi}{4})${}}
\newcommand{\cPhase}{controlled-\Phase{}}

\newcommand{\cells}[4]{
  \centering
  \Ket{
   \,
  \begin{tabular}{ | p{2.2mm} | p{2.2mm} | }
   \hline			
     #1 & #2   \\ \hline
     #3 & #4   \\ \hline 
   \end{tabular}\,
  } 
}

\newcommand{\barrier}{\cellcolor{orange}}

\newtheorem{Cl}{Claim}

\newtheorem{Def}{Definition}

\begin{document}

\title{A Quantum Game of Life}

\author[lab1]{P. Arrighi}{Pablo Arrighi}
\address[lab1]{Universit\'e de Grenoble, LIG}  
\email{pablo.arrighi@imag.fr}  
\urladdr{\href{http://membres-lig.imag.fr/arrighi/}{http://membres-lig.imag.fr/arrighi}}  

\author[lab2]{J. Grattage}{Jonathan Grattage}
\address[lab2]{Ecole Normale Sup\'erieure de Lyon, LIP}	
\email{jonathan.grattage@ens-lyon.fr}  
\urladdr{\href{http://www.grattage.co.uk/jon}{http://www.grattage.co.uk/jon}}  

\keywords{cellular automata, quantum computation, universality}

\begin{abstract}
This research describes a three dimensional quantum cellular automaton (QCA) which can simulate all other 3D QCA. This intrinsically universal QCA belongs to the simplest subclass of QCA: Partitioned QCA (PQCA). PQCA are QCA of a particular form, where incoming information is scattered by a fixed unitary $U$ before being redistributed and rescattered. Our construction is minimal amongst PQCA, having block size $2\times 2\times 2$ and cell dimension $2$. Signals, wires and gates emerge in an elegant fashion.
\end{abstract}

\maketitle

\section{Introduction}\label{sec:introduction}\label{subsec:CA}

\noindent{\em (Quantum) Cellular Automata.} Cellular automata (CA), first introduced by von Neumann \cite{Neumann}, consist of an array of identical cells, each of which may take one of a finite number of possible states. The entire array evolves in discrete time steps by iterating a function $G$. This global evolution $G$ is shift-invariant (it acts the same everywhere) and causal (information cannot be transmitted faster than some fixed number of cells per time step). Because this is a physics-like model of computation \cite{MargolusPhysics}, Feynman \cite{FeynmanQCA}, and later Margolus \cite{MargolusQCA}, suggested early in the development of quantum computation (QC) that quantising this model was important. This was for two main reasons. Firstly, in QCA computation occurs without extraneous (unnecessary) control, hence eliminating a source of decoherence, which is problem for QC implementation. Secondly, they are a good framework in which to study the quantum simulation of a quantum system, which is predicted to be a major application of QC. From a Computer Science perspective there are other reasons to study QCA, such as studying space-sensitive problems in computer science (\eg `machine self-reproduction', `Firing Squad Synchronisation', \ldots) in the quantum setting, or having more complete and formal models of physics ability to compute.  There is also the theoretical physics perspective, where CA are used as toy models of quantum space-time \cite{LloydQG}. The first approach to defining QCA \cite{ArrighiMFCS,DurrWell,Watrous} was later superseded by a more axiomatic approach \cite{ArrighiUCAUSAL,ArrighiLATA,SchumacherWerner} together with more operational approaches \cite{BrennenWilliams,NagajWocjan,PerezCheung,Raussendorf,VanDam,Watrous}.

\paragraph{\noindent{\em Intrinsic universality.}} Probably the most well known CA is Conway's `Game of Life', a two-dimensional CA which has been shown to be universal for computation, in the sense that any Turing Machine (TM) can be encoded within its initial state and then executed by evolution of the CA.  As TM are often regarded as a robust definition of `what an algorithm is' in classical computer science, this provides a key result in CA research. However, more can be achieved using CA than just running any algorithm. They run distributed algorithms in a distributed manner, model phenomena together with their spatial structure, and allow the use of the spatial parallelism inherent to the model. These features, modelled by CA and not by TM, are all worthy of investigation, and so the concept of universality should be revisited in this context to account for space. This is achieved by returning to the original meaning of the word \emph{universality} \cite{AlbertCulik,Banks,DurandRoka}, namely the ability for one instance of a computational model to be able to simulate other instances of the same computational model. Intrinsic universality formalises the ability of a CA to simulate another in a space-preserving manner \cite{MazoyerRapaport,OllingerJAC,Theyssier}, and has been extended to the quantum setting \cite{ArrighiUQCA,ArrighiPQCA,ArrighiSimple}. In previous work \cite{ArrighiSimple} it was shown that, when the dimension of space is greater than one, the problem of intrinsic universality reduces to the ability to code for signals, wires and a universal set of quantum gates, as expected from the classical CA case \cite{OllingerJAC}.

\paragraph{\noindent{\em Related work.}} There are several related results in the current CA literature. For example, several works \cite{MargolusPhysics,MoritaCompUniv1D,MoritaCompUniv2D} provide computation universal Reversible Partitioned CA constructions, and \cite{MoritaIntrinsicUniv1D} deals with their ability to simulate any CA in the one-dimensional case. The problem of minimal intrinsically universal  CA  has been addressed \cite{OllingerRichard}, and for Reversible CA (RCA) the issue of intrinsic universality has been tackled \cite{Durand-LoseLATIN,Durand-LoseIntrinsic1D,FredkinToffoli}. The difficulty is in having an $n$-dimensional RCA simulate all other $n$-dimensional RCA and not, say, the $(n-1)$-dimensional RCA, otherwise a history-keeping dimension could be used, as shown by Toffoli \cite{ToffoliConstruction}. Concerning QCA, Watrous \cite{WatrousFOCS} proved that QCA are universal in the sense of QTM. Shepherd, Franz and Werner \cite{ShepherdFranz} defined a class of QCA where the scattering unitary $U_i$ changes at each step $i$ (CCQCA). Universality in the circuit-sense has been achieved by Van Dam \cite{VanDam}, Cirac and Vollbrecht \cite{VollbrechtCirac}, Nagaj and Wocjan \cite{NagajWocjan} and Raussendorf \cite{Raussendorf}. In the bounded-size configurations case, circuit universality coincides with intrinsic universality, as noted by Van Dam \cite{VanDam}. QCA intrinsic universality in the one-dimensional case has been resolved \cite{ArrighiFI}, and also in the general $n>1$-dimensional case \cite{ArrighiSimple}. Both results rely on recent work \cite{ArrighiPQCA}, where it was shown that a simple subclass of QCA, namely Partitioned QCA (PQCA), are intrinsically universal. This enables the focus here to be on this simple, natural class of QCA, as previously proposed \cite{BrennenWilliams,Karafyllidis,NagajWocjan,Raussendorf,SchumacherWerner,VanDam,Watrous}.  PQCA are QCA of a particular form, where incoming information is scattered by a fixed unitary $U$ before being redistributed and rescattered. Hence the problem of finding an intrinsically universal PQCA is reduced to finding some scattering unitary $U$ that supports the implementation of signals, wires and a universal set of quantum gates.

\paragraph{\noindent{\em Game of Life.}} As the more general case of finding a general $n$-dimensional intrinsically universal QCA has been resolved \cite{ArrighiPQCA}, one could question the necessity of focusing on the particular 3D case of this problem. Our answer to this question is threefold. 
\begin{itemize}
\item In the study of models of computation, the definition of the model is often followed by the search for a universal instance, which is then followed by a search for a minimal universal instance. Reaching this final step is generally regarded as a sign of maturity of the model. It shows that the entire model can be reduced to a particular simplest instance. Moreover, sometimes a particular instance stands out, either for its elegance and simplicity, or the richness of its behaviour. This was clearly the case with the Game of Life within the realm of Classical CA, and it is echoing this impression that we have named this QCA the `Quantum Game of Life'. This is not because of a resemblance in the construction of the local rule, which clearly owes more to the Billiard Ball Model CA \cite{FredkinToffoli}.
\item Dimension three is particularly interesting, not only because of the relevance to physics, but because intrinsic universality QCA constructions lend themselves to a striking simplification. In dimension one the simplest known intrinsically universal QCA has cell dimension 36 \cite{ArrighiFI}. In dimension two it has cell dimension $4$ \cite{ArrighiSimple}, and seems reducible to $3$ via a costly and inelegant variation of the scheme \cite{ArrighiSimple}, but probably no further. The solution we present here is in dimension three and has cell dimension $2$; it is therefore minimal.
\item Animations of the QCA operating, plus an implementation of this QCA, are available on the companion website \cite{Website3D}. Although it lacks a friendly way of preparing the initial configuration, it turns out to be quite interesting and entertaining already, as is the Game of Life. We plan to implement the QCA in a more graphical, interactive, computationally efficient fashion. The end result could be a pedagogical tool, both for discussing quantum theory and in order to start exploring the dynamical behaviour of QCA, such as in \cite{NesmeGutschow}, perhaps using methods inspired by similar work on Probabilistic CA \cite{RST}. 
\end{itemize}

\section{The Universal QCA}\label{sec:theuniv}

\paragraph{\em Definitions.}
Configurations hold the basic states of an  entire array of cells, and hence denote the possible basic states of the entire QCA: 
\begin{Def}[Finite configurations]
A \emph{(finite) configuration} $c$ over $\Sigma$ is a function $c:
\Z^3 \longrightarrow \Sigma$, with $(i_1,i_2,i_3)\longmapsto
c(i_1,i_2,i_3)=c_{i_1,i_2,i_3}$, such that there exists a (possibly empty)
finite set $I$ satisfying $(i_1,i_2,i_3)\notin I\Rightarrow c_{i_1,i_2,i_3}=q$, where $q$ is a distinguished \emph{quiescent} state of $\Sigma$.
The set of all finite configurations over $\Sigma$ will be denoted $\mathcal{C}^{\Sigma}_f$.
\end{Def}
As this work relates to QCA rather than classical CA, the global state can be a superposition of these configurations. To construct the separable Hilbert space of superpositions of configurations the set of configurations must be countable. This is why finite, unbounded, configurations are considered; the quiescent state of a CA is analogous to the blank symbol of a TM tape. 
\begin{Def}[Superpositions of configurations]\label{superp} 
Let $\mathcal{H}_{\mathcal{C}^{\Sigma}_f}$ be the Hilbert space of configurations. Each finite configuration $c$ is associated with a unit vector $\ket{c}$, such that the family $\pa{\ket{c}}_{c\in\mathcal{C}^{\Sigma}_f}$ is an orthonormal basis of $\mathcal{H}_{\mathcal{C}^{\Sigma}_f}$. A \emph{superposition of 
configurations} is then a unit vector in $\mathcal{H}_{\mathcal{C}^{\Sigma}_f}$. 
\end{Def}
Detailed explanations of these definitions, as well as axiomatic definitions of QCA, are available \cite{ArrighiUCAUSAL,ArrighiLATA,SchumacherWerner}. Building upon these works, we have shown \cite{ArrighiPQCA,ArrighiSimple} that Partitioned QCA (PQCA) are intrinsically universal. Since they are the most canonical description of QCA, and since the aim of this paper is to construct the most canonical, and yet universal, QCA, we will assume that all QCA are PQCA throughout this work.
\begin{Def}[Partitioned QCA]\label{def:pqca}
A partitioned  three-dimensional quantum cellular automaton (PQCA) is defined by a \emph{scattering unitary}, a unitary operator, $U$ such that $U:\mathcal{H}_{\Sigma}^{\otimes 2^3}\longrightarrow\mathcal{H}_{\Sigma}^{\otimes 2^3}$, and $U\ket{qq\ldots qq}=\ket{qq\ldots qq}$, \ie that takes
a cube of $2^3$ cells into a cube of $2^3$ cells and preserves quiescence. Consider $G=(\bigotimes_{2\mathbb{Z}^3} U)$, the operator over $\mathcal{H}$. The induced global evolution is $G$ at odd time steps, and $\sigma G$ at even time steps, where $\sigma$ is a translation by one unit in all directions.
\end{Def}
Here we provide a specific instance of a $U$-defined PQCA which is capable of intrinsically simulating any $V$-defined PQCA, for any $V$. In order to describe such a $U$-defined PQCA in detail, two things are required: the dimensionality of the cells (including the meaning attached to each of the states they may take); and the way in which the scattering unitary $U$ acts upon these cells.

\subsection{Signals and wires}

Classical CA studies often refer to `signals' without an explicit definition. In this context, a signal refers to the state of a cell which may move to a neighbouring cell consistently, from one step to another, by the evolution of the CA. Therefore a signal would appear as a line in the space-time diagram of the CA.  

The $U$-defined PQCA constructed in this paper is minimal, in the sense that it is a 3D PQCA with the smallest non-trivial block-size and cell-dimension. Hence, each cell has only two possible states, either empty or occupied; it is a binary automaton, with scattering unitary $U$ acting on $2\times 2 \times 2$ cell neighbourhoods (a cube). The description of the scattering unitary $U$ that defines the behaviour of the automaton will now be provided in a case by case manner.

\subsubsection{Signals.}

\begin{figure}
\centering
\includegraphics[scale=.30, clip=true]{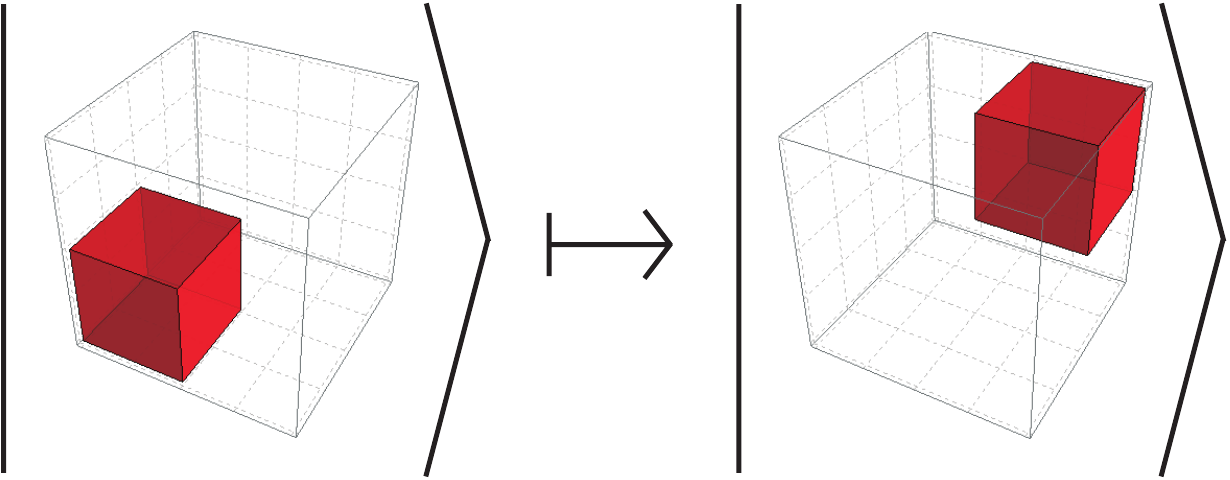}
\caption{Signals propagate diagonally across the partition.}
\label{fig:move}
\end{figure}

Signals in our scheme will be encoded as isolated occupied cells.  
If a neighbourhood trivially contains no signals, then no change is made. 
If a neighbourhood contains only one signal, \ie only one cell is occupied, and the other seven are empty, then the signal is shifted by one in every direction, yielding a 3D diagonal propagation. This rule is given in Fig.~\ref{fig:move}.
These two rules, together with their rotations, define the action of the scattering unitary $U$ upon the subspace with zero or one occupied cells. Notice that $U$ can easily be seen to be unitary upon this subspace, as it performs a permutation of the basis states. 

\subsubsection{Barriers.}\label{sec:barriers}

\begin{figure}
\centering
\includegraphics[scale=.11, clip=true]{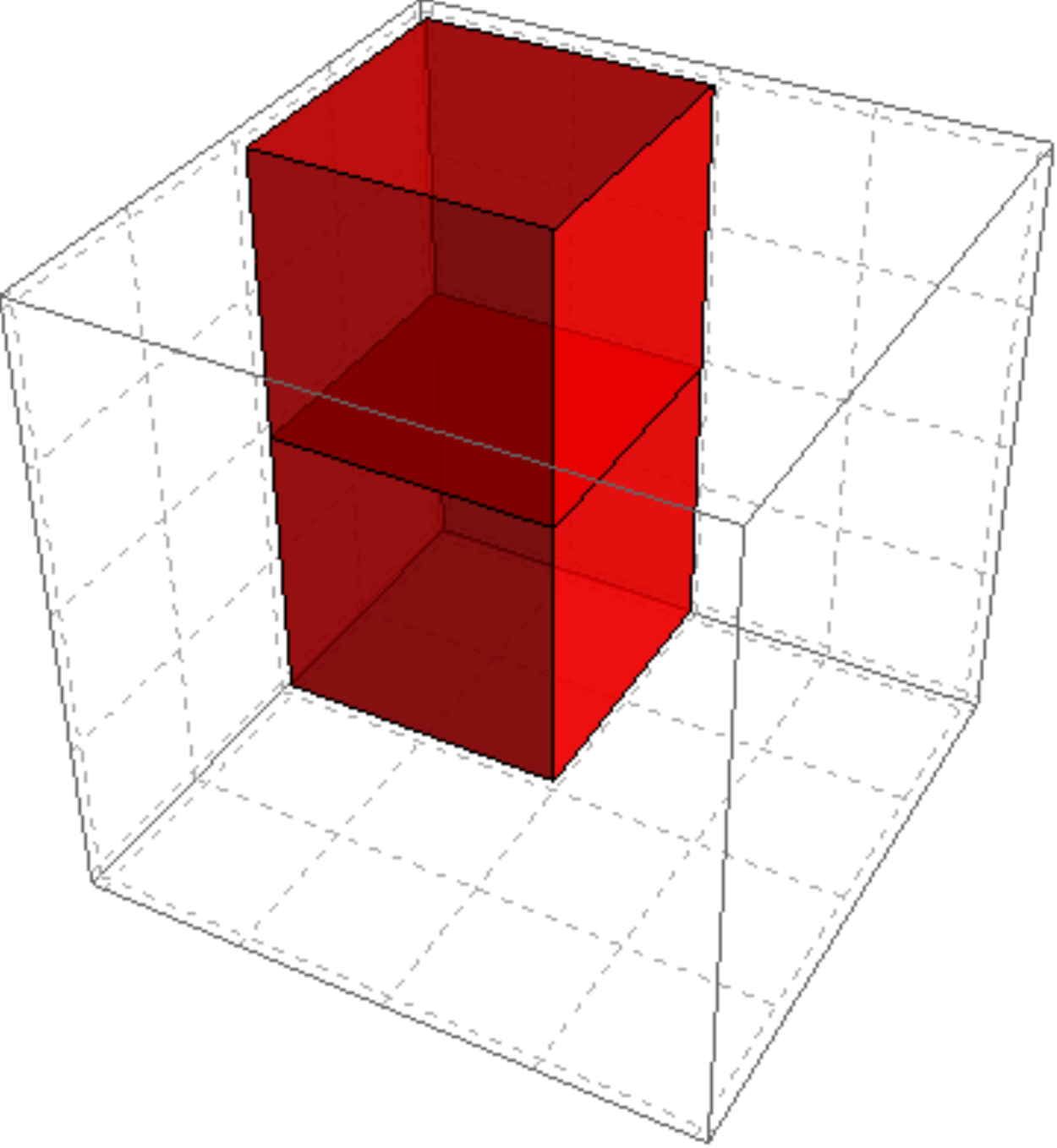}
\caption{Two occupied cells which share one face form a static barrier.}
\label{fig:barrier}
\end{figure}

\begin{figure}
\centering
\includegraphics[scale=.15, clip=true]{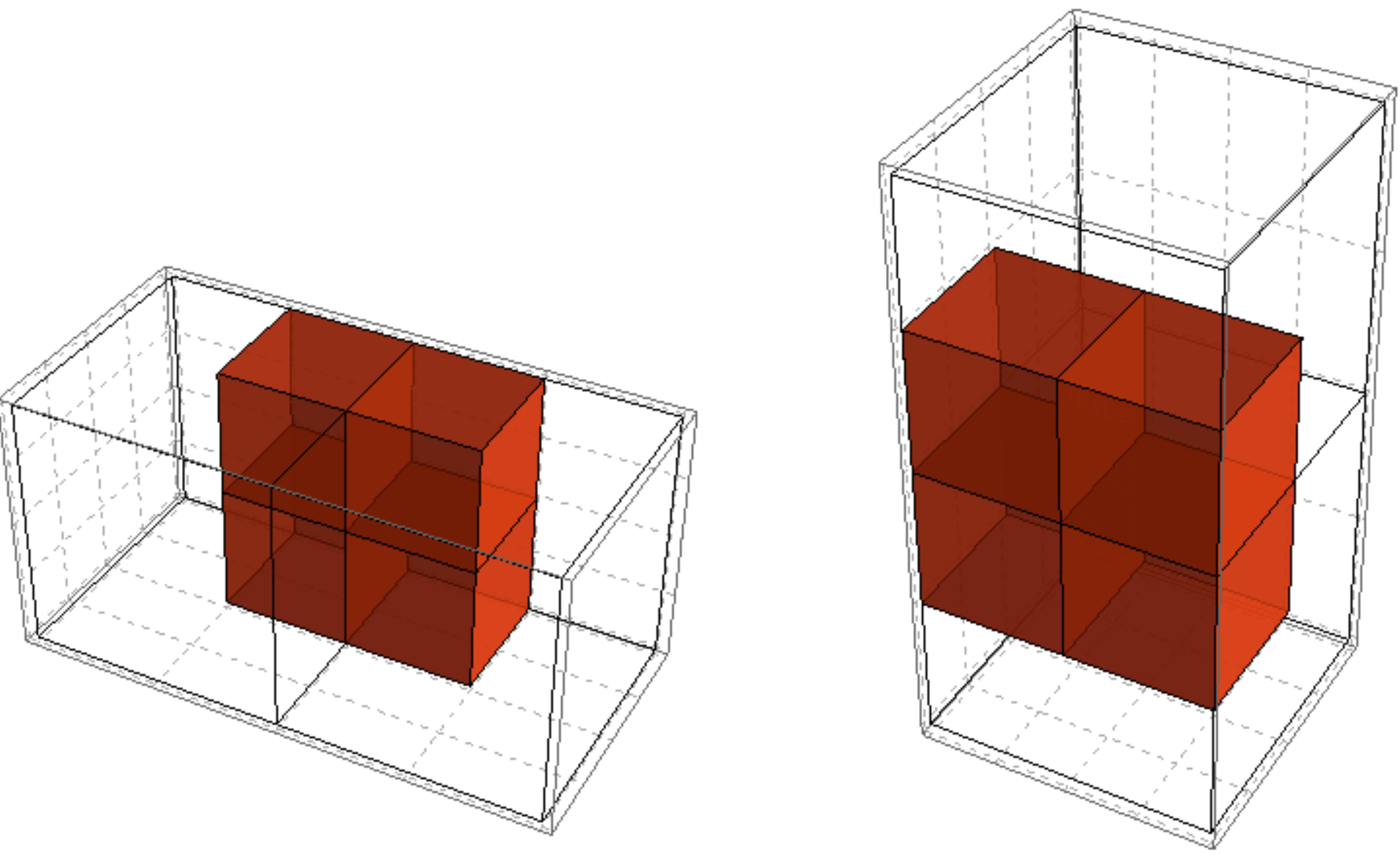}
\caption{Two vertical barriers placed in adjacent partitions horizontally form a $2\times 2\times 1$ barrier (\emph{left}). At the next time step, the repartitioning is still stable, with two horizontal barriers in adjacent vertical partitions formed from those same cells (\emph{right}).}
\label{fig:twobarriers}
\end{figure}

Two or more occupied cells that share a common face form a barrier, which is intended to be a stationary pattern. The rule is given in Fig.~\ref{fig:barrier}. 
Such a barrier is stable for one partition, but in the next partition it appears as two signals moving away from each other, and hence it will scatter. This can be avoided by extending beyond the current partition, using at least four occupied cells. For example, a $2 \times 2 \times 1$ block of cells which overlaps a partition and forms a barrier in each partition, as shown in Fig.~\ref{fig:twobarriers}, is stable.
This rule, together with its rotations, defines the action of the scattering unitary $U$ upon the subspace of two adjacent occupied cells. Upon this subspace the scattering unitary $U$ acts like the identity, which is unitary. 

\subsubsection{Walls.}

\begin{figure}
\centering
\includegraphics[scale=.10, clip=true]{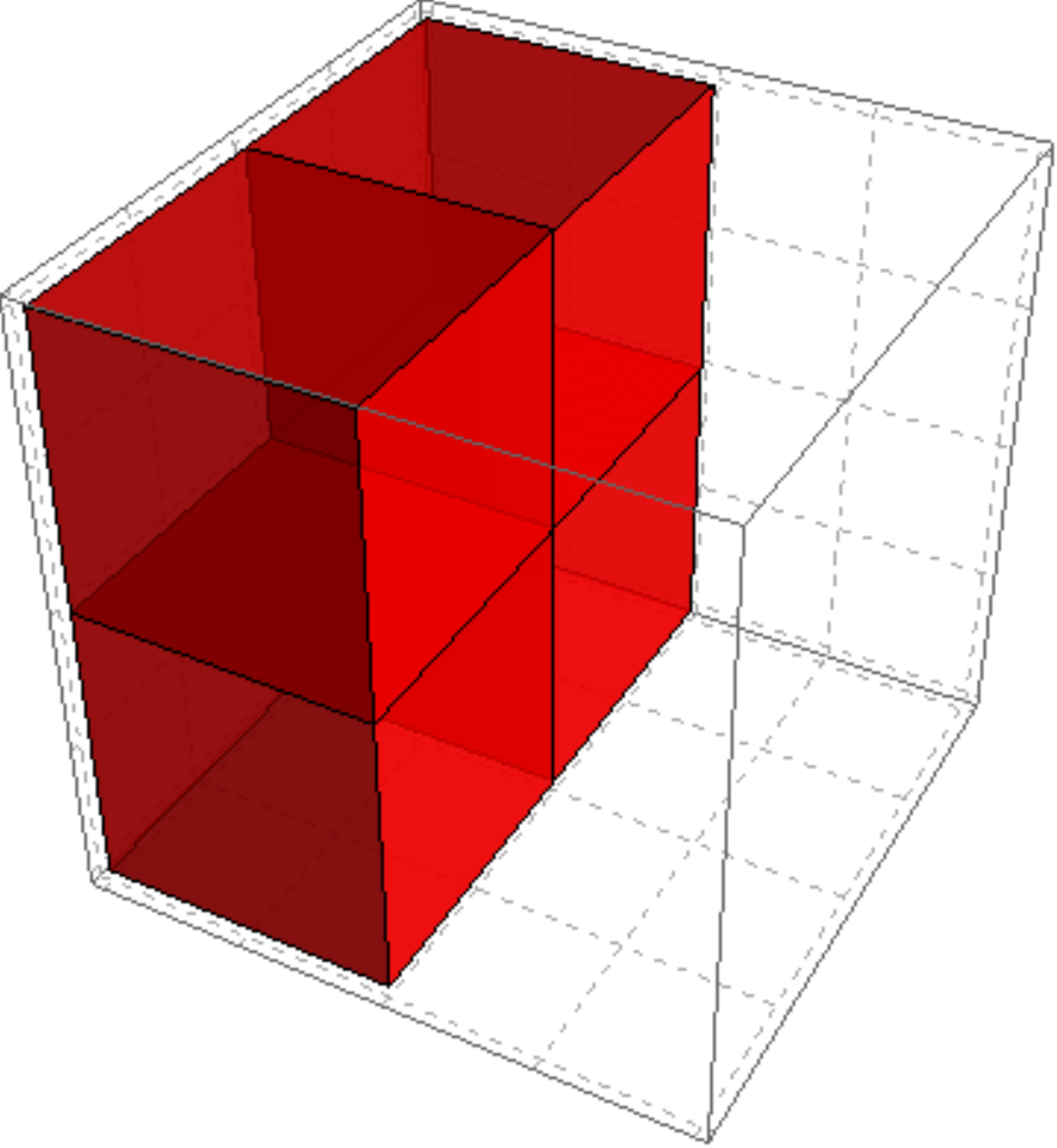}
\caption{Four occupied cells occupying a face of the partition neighbourhood form a wall.}
\label{fig:barrierWall}
\end{figure}

\begin{figure}
\centering
\includegraphics[scale=.30, clip=true]{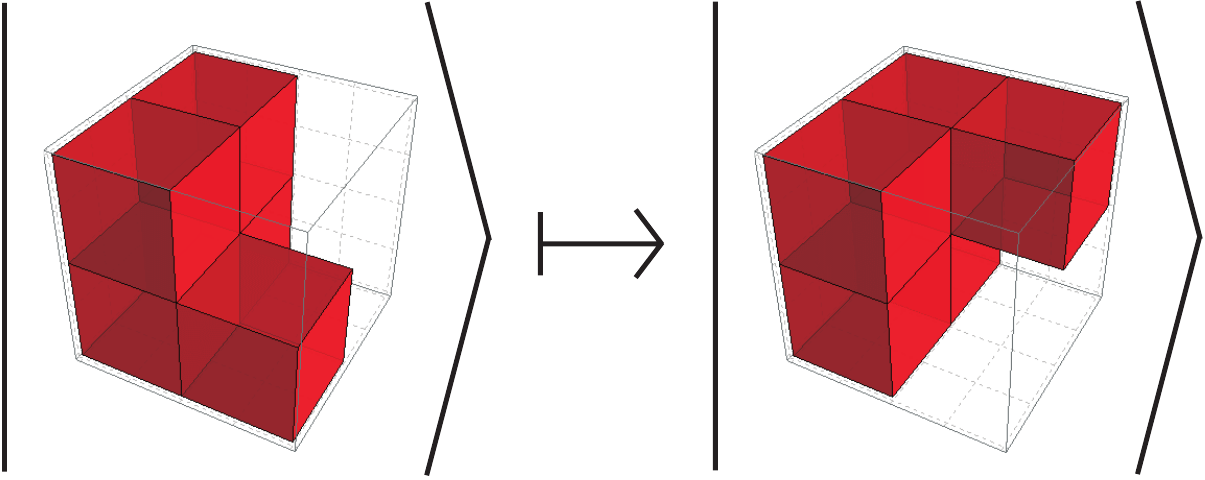}
\caption{A signal which interacts with a wall in a neighbourhood ``bounces'' off the wall, causing it to change direction along one axis. Compare with the unhindered signal propagation shown in Fig.~\ref{fig:move}.}
\label{fig:bounce}
\end{figure}

When two barriers, made of four occupied cells, form a square that fits within a partition, as in Fig.~\ref{fig:barrierWall}, we say that they form a wall. A wall is able to redirect a fifth occupied cell, that shares only one face with the wall, which represents the incoming signal.
Again a wall in one partition appears as signals moving away from each other in the next partition, but again they can be made stable following Fig.~\ref{fig:twobarriers}, by extending them across partitions.
The behaviour required is for an incoming signal to ``bounce'' off the wall, and the rule producing this behaviour is given in Fig.~\ref{fig:bounce}. This rule suffices to define signal redirection, or ``rewiring''.
These rules, together with their rotations, define the action of the scattering unitary $U$ upon the subspace of walls and walls with a signal. Notice that $U$ is unitary upon this subspace, as it performs a permutation of the basis states.

\subsection{Qubits and quantum gates}

\begin{figure}
\centering
\includegraphics[scale=.40, clip=true]{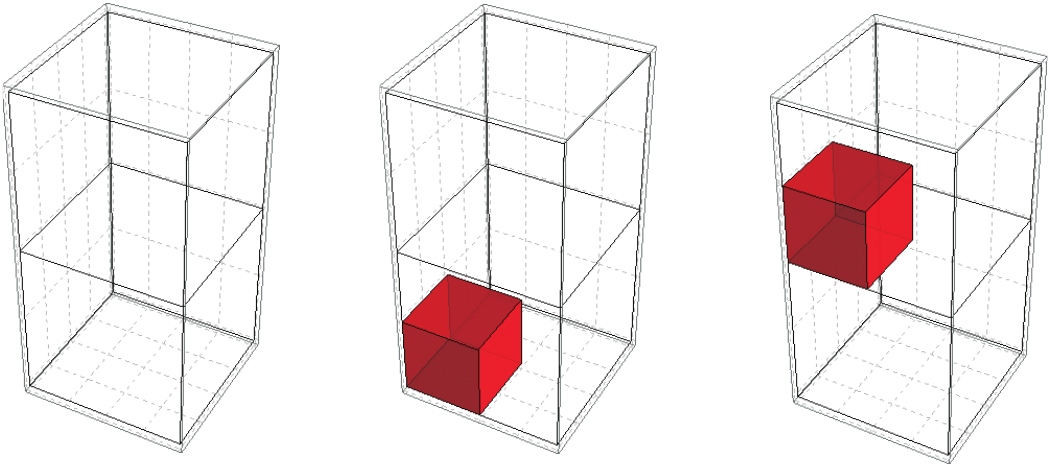}
\caption{Qubits are implemented as parallel tracks of signals. No qubit is modelled as no signal (\emph{left}), $\ket{0}$ is modelled as a signal on the lower track (\emph{centre}), while a signal on the upper track models a qubit in state $\ket{1}$ (\emph{right}).}
\label{fig:signals}
\end{figure}

To allow a universal set of gates to be implemented by the PQCA, certain combinations of signals and barriers need to be given special importance. 
To implement qubits, pairs of signals will be used. Indeed in our scheme a qubit is formed by two parallel ``tracks'' of signals, as shown in Fig.~\ref{fig:signals}. A signal on the bottom track indicates a $\ket{0}$ state, whereas the top track indicates a $\ket{1}$ state. Qubits in superpositions are modelled by appropriate superpositions of the configurations, as in Def.~\ref{superp}. We can now define rules which implement quantum gates on those qubits\footnote{In Quantum Mechanics, a qubit can be encoded into any 2 degrees of freedom of a system. Some experiments in quantum optics, for example, use the spin degree of freedom, whereas others use a spatial degree of freedom. This model is analogous to the latter case. See the conclusion section for a discussion of relevance to physical systems.}. 

\subsubsection{Hadamard.}
\begin{figure}
\centering
\includegraphics[scale=.35, clip=true]{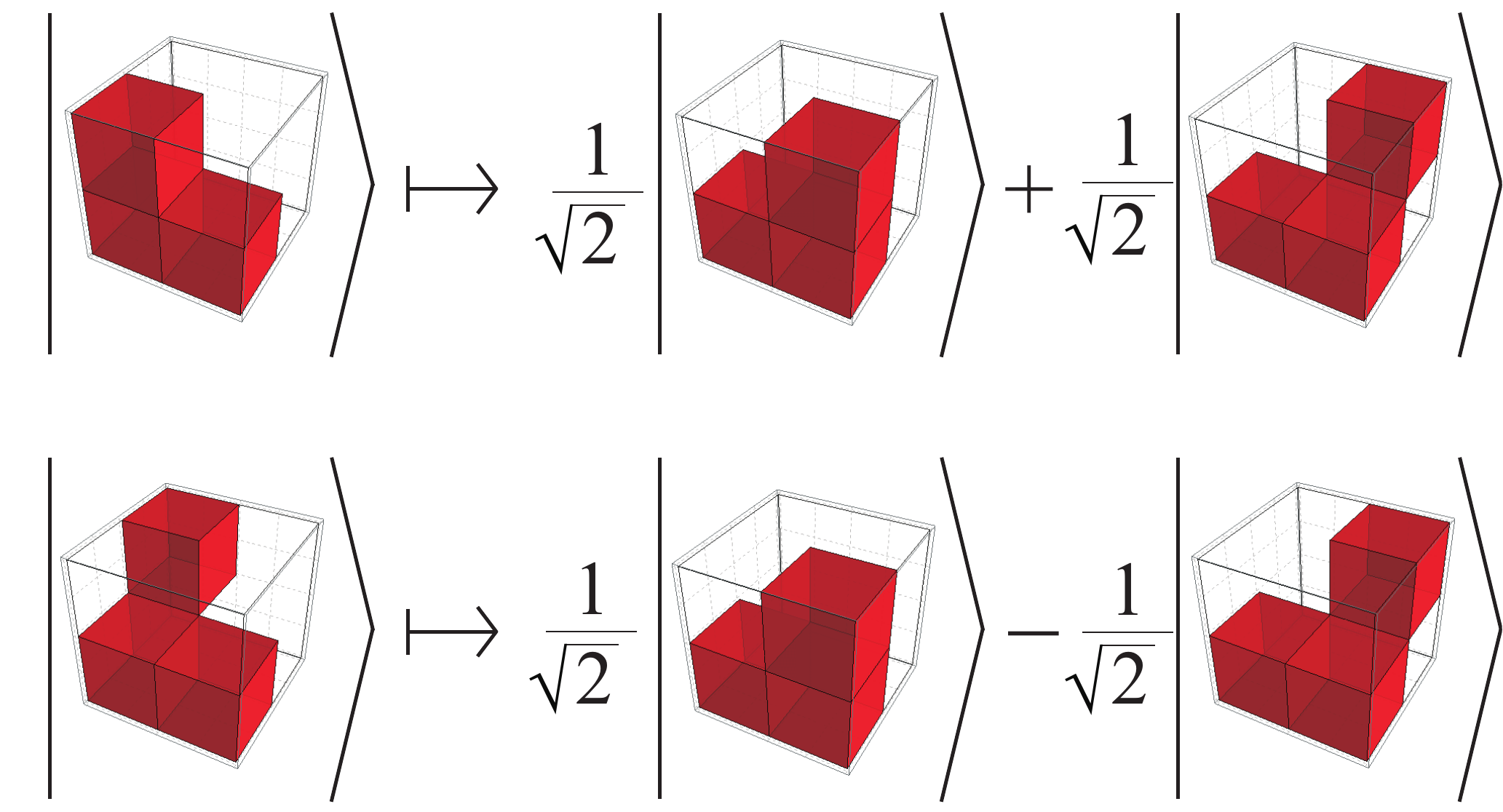}
\caption{The rules implementing the Hadamard operation. The first gives $\Ket{0} \mapsto \frac{1}{\sqrt{2}}\Ket{0}+\frac{1}{\sqrt{2}}\Ket{1}$ while the second gives $\Ket{1} \mapsto \frac{1}{\sqrt{2}}\Ket{0}-\frac{1}{\sqrt{2}}\Ket{1}$.}
\label{fig:had}
\end{figure}

The Hadamard gate is a requirement for universal quantum computation, where $H: \ket{0} \mapsto \frac{1}{\sqrt{2}}\ket{0}+\frac{1}{\sqrt{2}}\ket{1}, \ket{1} \mapsto \frac{1}{\sqrt{2}}\ket{0}-\frac{1}{\sqrt{2}}\ket{1}$. In order to achieve this typically quantum behaviour, a special meaning is attached to the interaction of a signal with a single two-cell barrier (as defined in section \ref{sec:barriers}). The corresponding rule is given by Fig.~\ref{fig:had}.

As all the rules we provide are rotation invariant, the unambiguity of the Hadamard rule may not be clear. 
To be  unambiguous, the $\ket{0}$ and $\ket{1}$ cases need be to distinguished from each other, and the signal distinguishable from the barrier.
This can be done as follows: if the cube can be rotated such that all three signals are on a single face, forming an {``L''} shape, then the input to the Hadamard is $\ket{0}$, and the signal is at the top of the L. If instead a lone occupied cell is on the top left of the furthest face, with a two-cell barrier along the bottom of the closest face, forming a dislocated  {L} shape, then the input to the Hadamard is $\ket{1}$ and the signal is the isolated, top left cell.

\begin{figure}
\centering
\includegraphics[scale=.50, clip=true]{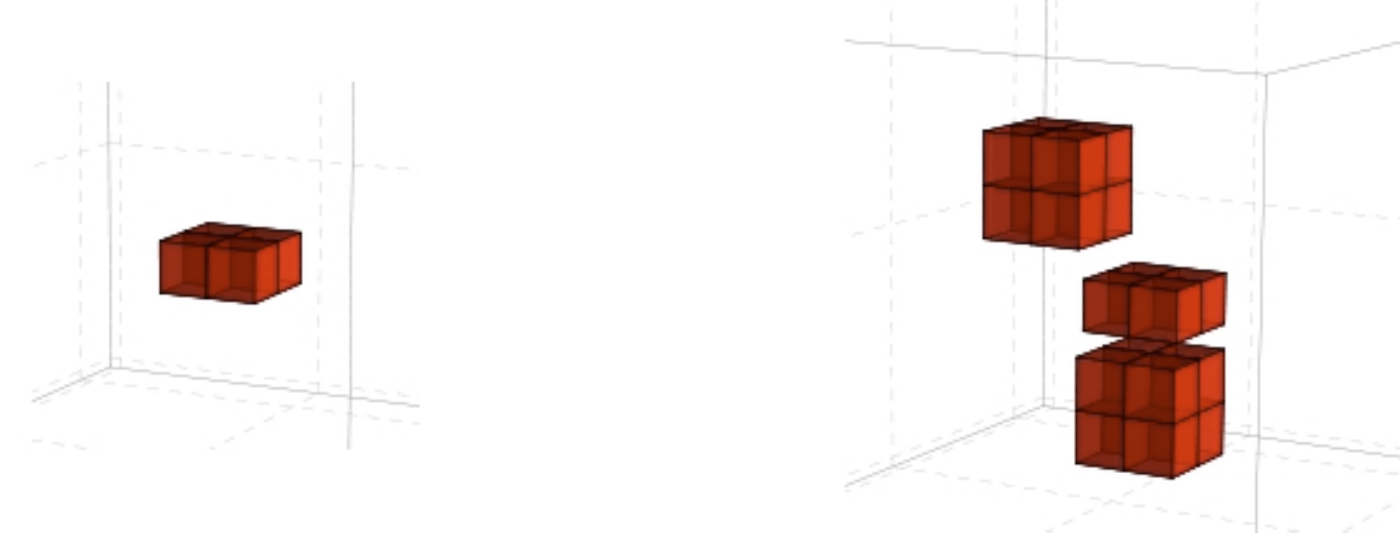}
\caption{A stable Hadamard configuration, formed by two barriers extended across partitions (\emph{left}). The Hadamard configuration can be extended with two cubed ``walls''. The first cube, on the top left, redirects the top track $\ket{1}$ signal, incoming from the bottom left and travelling to the far top right, into the barrier. The second cube, bottom right, redirects the output bottom $\ket{0}$ track so that it is again moving parallel to the $\ket{1}$ track, keeping the qubit representation consistent (\emph{right}).}
\label{fig:hadGate}
\end{figure}

On its own, a barrier is not enough to implement the Hadamard gate, because it does not respect the parallel track convention of qubits defined earlier. Rather, inputs and outputs are orthogonal. To rectify this, the barrier can be
combined with two walls, which redirect the signals appropriately, as shown in Fig.~\ref{fig:hadGate}. Further rewiring can be added so the qubit can arrive and leave in any direction, but these are not easily represented in 2D projections.

These two rules, together with their rotations, define the action of the scattering unitary $U$ upon the subspace of two-cell barriers plus a signal. Notice that $U$ is unitary upon this subspace, as it is a composition of a Hadamard gate upon the input states, together with the permutation of the basis states that perform the movement described in Fig.~\ref{fig:had}. 

\subsubsection{Controlled rotations.}

\begin{figure}
\centering
\includegraphics[scale=.30, clip=true]{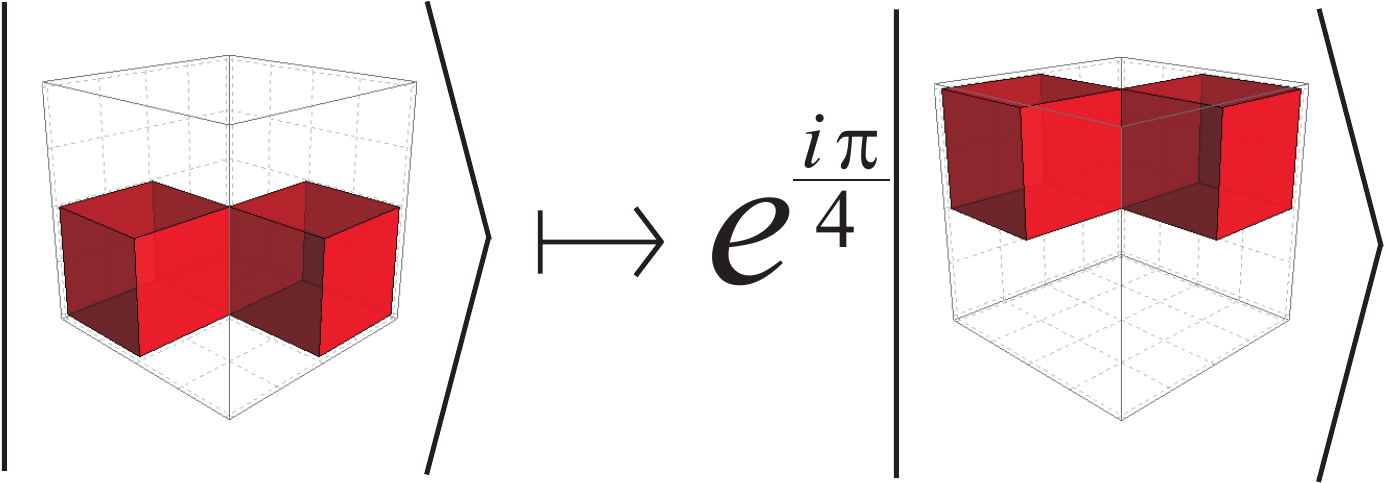}
\caption{When two signals cross each other diagonally, a complex phase is added to the configuration state.}
\label{fig:phase}
\end{figure}

A two qubit controlled gate is another requirement for universal quantum computation, and in this case we choose the \cPhase gate, where \textsc{cR(${\pi}\slash{4}$)}: $\ket{11} \mapsto  e^{\frac{i\pi}{4}}\ket{11}$, and is the identity otherwise. To encode a two qubit controlled gate, signal collisions, where two moving signals cross each other diagonally, will be interpreted as adding a global phase of $e^{\frac{i\pi}{4}}$. This rule is given by Fig.~\ref{fig:phase}.

This allows a \cPhase{} operation to be defined, by redirecting the $\ket{1}$ (true) signal track so that the tracks will cross each other in this way, and then recreating the qubits by rejoining the $\ket{1}$ tracks back with the appropriate $\ket{0}$ tracks, using walls to redirect and delay signals as in the Hadamard case. Again, this 3D configuration cannot easily be presented in 2D.

This same rule can also be used to implement a single qubit phase change operation by causing a control signal to loop such that it will intersect with the true track of the qubit, thus adding the global phase to the configuration. This needs to be correctly timed so that the tile implementing the single qubit rotation can be iterated, following the same reasoning as in \cite{ArrighiSimple}.

This rule defines, together with its rotations, the action of the scattering unitary $U$ upon the subspace of two non-adjacent signals on face. $U$ can again be seen to be unitary upon this subspace, as it is a permutation of the basis states, with a phase.

\subsection{Universality, erosion and dynamics}

We have seen that the scattering unitary $U$ is a unitary on the subspaces upon which it has been defined. Additionally, following arguments given previously \cite{ArrighiSimple}, it may be concluded that the $U$-defined PQCA given here is intrinsically universal:
\begin{itemize}
\item In space dimension greater than one, it suffices to implement signals, wires and a universal set of gates to achieve intrinsic universality; 
\item The \cPhase{} and the Hadamard gate are universal for Quantum Computation.  
\end{itemize}
Indeed the standard set of \textsc{cNot}, \textsc{H}, \textsc{\Phase} can be recovered as follows:
$$\textsc{cNot}\ket{\psi}=(\mathbb{I}\otimes H)(\textsc{cR(${\pi}\slash{4}$)})^4(\mathbb{I}\otimes H)\ket{\psi}$$
where $\textsc{cR($\frac{\pi}{4}$})^4$ denotes four applications of the \cPhase{} gate, giving the controlled-\textsc{Phase} operation.

We could let the action of the scattering unitary $U$ be the identity in all other subspaces. However, some patterns would then be unmovable and indestructible, such as the wall of Fig.~\ref{fig:barrierWall}. This would arguably make the dynamics of the QCA presented here a little plain. Moreover, since  walls could be built through simultaneous signal collision,  it would be natural to be able to dismantle them. If we are interested in the dynamics, different ways of completing the definition of $U$ upon the remaining subspaces should be considered.
A natural way is to consider that all of the other cases are made of non-interacting signals, where the occupied cells propagate past each other diagonally. This allows, for instance, for the demolition of walls and the creation of new stable patterns. 

Animated examples and an implementation of this cellular automaton can be found online \cite{Website3D}. We hope to explore its dynamics in future simulations.

\section{Comparison with Two-Dimensional Rules \& Tiles}

In previous work we provided a generic construction of an $n$-dimensional intrinsically universal QCA \cite{ArrighiSimple}. In this section we will compare the two dimensional instance given previously with the scheme presented here.

{\em Objects.} In \cite{ArrighiSimple} the cell-dimension is $4$. A cell can be either empty, a wall, or a binary signal ($0$ and $1$). In a sense the same objects are recovered in the scheme presented here, but as the cell dimension is $2$ these are conveyed as patterns of cells. Walls are now coded as $2^2$ squares as in Fig.~\ref{fig:barrierWall}. Signals no longer have a bit attached to them; instead the location of the signal is used to encode this bit of information. 

{\em Rules.} Once the above observations have been made; it can be seen that the two dimensional, cell-dimension four scheme previously described \cite{ArrighiSimple} is almost a projection of the scheme presented here. Indeed:\\
- the 2D signal travelling rule of \cite{ArrighiSimple};  $\cells{}{}{$s$}{} \mapsto \cells{}{$s$}{}{}$, is a height-projection of the corresponding rule in Fig.~\ref{fig:move};\\
- the 2D signal bouncing rule;  $\cells{\barrier}{\barrier}{s}{} \mapsto \cells{\barrier}{\barrier}{}{$s$}$, again a height-projection of the corresponding rule in Fig.~\ref{fig:bounce};\\
- the 2D control-phase rule;  
$$\cells{1}{}{1}{} \mapsto e^{\frac{i\pi}{4}}\cells{}{1}{}{1}, \quad \cells{x}{}{y}{} \mapsto \cells{}{y}{}{x} otherwise,$$
is a width-depth projection of the corresponding rule in Fig.~\ref{fig:phase};\\
- only the 2D Hadamard rule of \cite{ArrighiSimple} differs slightly:
$$\cells{\barrier}{}{0}{\barrier} \mapsto\frac{1}{\sqrt{2}}\cells{\barrier}{0}{}{\barrier} + \frac{1}{\sqrt{2}}\cells{\barrier}{1}{}{\barrier}$$
$$\cells{\barrier}{}{1}{\barrier} \mapsto \frac{1}{\sqrt{2}}\cells{\barrier}{0}{}{\barrier} - \frac{1}{\sqrt{2}}\cells{\barrier}{1}{}{\barrier}$$

\begin{figure}
\centering
\VS{\vspace{-0.5mm}\includegraphics[scale=.55, clip=true]{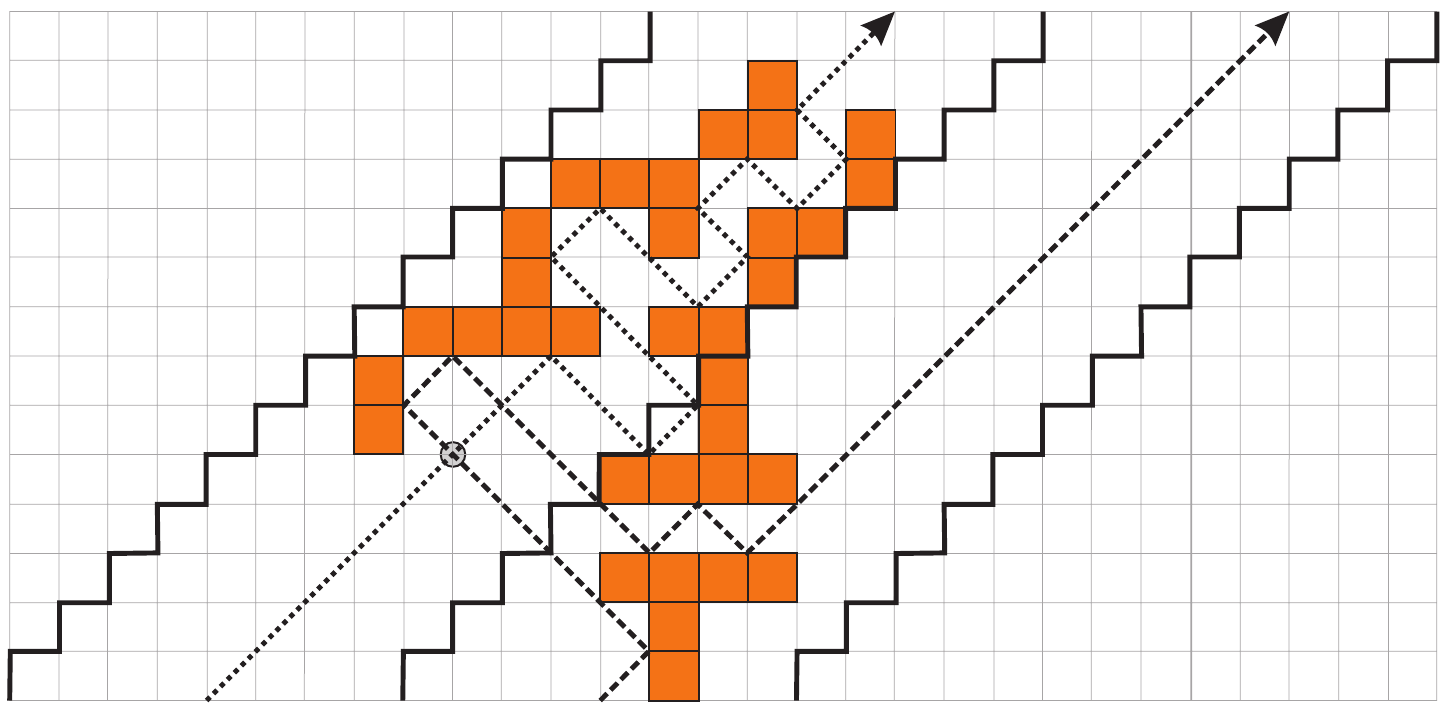}}
\VL{\includegraphics[scale=.60, clip=true]{images/cPhaseCirc.pdf}}
\caption{The $2$D `\cPhase{} gate' tile\VS{, with a signal interaction at the highlighted cell.}\VL{ applies the controlled-\Phase{} operation to the two input qubits, by causing the signals to interact at the highlighted point (grey circle). The qubits are then synchronised so that they exit at the same time along their original paths. No swapping takes place.}}
\label{fig:cPhaseTile}
\end{figure}

{\em Tiles.} Previously,  not only were the above rules provided, but a set of fixed-sized ($16 \times 14$) tiles, taking a fixed number of time steps ($24$) to be travelled through, and each accomplishing one of the universal quantum gates upon the incoming signals were also given \cite{ArrighiSimple}. The example of the \cPhase{} tile is reproduced in Fig.~\ref{fig:cPhaseTile} for concreteness. The difference between such tiles and the corresponding rule (the \cPhase{} rule in this case) is that tiles need only to be placed next to one another so as to make up a circuit, with all signal wiring and synchronisations being accounted for.
It would be cumbersome to describe the equivalent of all these tiles in the scheme presented here, particularly due to their three dimensional nature. Some are shown on the companion website \cite{Website3D}. The reader may convince himself of their feasibility through the following arguments: since those tiles exist in two-dimensions, and since the two-dimensional rules are projections of the three-dimensional rules, there must exist some three-dimensional tiles for which the two-dimensional ones are, in some sense, projections.

\section{Conclusion}\label{sec:discussion}

This paper presents a minimal $3$D PQCA which is capable of simulating all other PQCA, preserving the topology of the simulated PQCA. This means that the initial configuration and the forward evolution of any PQCA can be encoded within the initial configuration of this PQCA, with each simulated cell encoded as a group of adjacent cells in the PQCA, \ie intrinsic simulation. The main, formal result of this work can therefore be stated as: 
\begin{Cl}
There exists an $3$D $U$-defined PQCA, with block size $2$ and cell dimension $2$, which is an intrinsically universal PQCA. Let $H$ be a $3$-dimensional $V$-defined PQCA such that $V$ can be expressed as a quantum circuit $C$ made of gates from the set $\textsc{Hadamard}$, $\textsc{Cnot}$, and $\textsc{\Phase}$. Then $G$ is able to intrinsically simulate $H$.
\end{Cl}
Any finite-dimensional unitary $V$ can always be approximated by a circuit $C(V)$ with an arbitrary small error $\varepsilon=\max_{\ket{\psi}}||V\ket{\psi}-C\ket{\psi}||$. Assuming instead that $G$ simulates the $C(V)$-defined PQCA, for a region of $s$ cells over a period $t$, the error with respect to the $V$-defined PQCA will be bounded by $st\varepsilon$. This is due to the general statement that errors in quantum circuits increase, at most proportionally with time and space \cite{NielsenChuang}.
Combined with the fact that PQCA are universal \cite{ArrighiSimple,ArrighiPQCA}, this means that $G$ is intrinsically universal, up to this unavoidable approximation.

{\em Discussion.}
This PQCA is definitely minimal amongst the $3$D PQCA. Reducing the block size or the cell dimension necessarily results in a trivial PQCA. Moreover, PQCA are the simplest and most natural class of QCA \cite{ArrighiHDR}. Nevertheless, it is not so clear that this PQCA is minimal amongst all intrinsically universal $3$D QCA. Indeed, PQCA-cells are really QCA-subcells, and PQCA-blocks need to be composed with them shifted in order to yield the QCA neighbourhood. Based on these observations our QCA has QCA-cell dimension $2^8$ and the $3$D radius--$\frac{1}{2}$ neighbourhood. Whether there are some further simplifications in this more general setting is an open question.


An important source of inspiration, along with the original Game of Life, was the the simplest known intrinsically universal classical Partitioned CA \cite{MargolusQCA}, which has cell dimension $2$. Called the BBM CA, it was itself directly inspired by the Billiard Ball Model \cite{FredkinToffoli}.
Our scheme also features signals (billiard balls, or particles) bouncing and colliding. This analogy with physics is a desirable feature; Feynman's sole purpose for the invention of QC \cite{FeynmanQC} was that quantum computers would be able to simulate quantum systems efficiently, and hence predict their behaviour. It is still thought that quantum simulation will be one of most important uses of quantum computers for society, with expected and potentially unexpected impact in quantum chemistry, biochemistry or nanotechnologies (\eg in terms of the synthesis of specific purpose molecules). This was also one of the reasons why Feynman immediately promoted the QCA model of QC \cite{FeynmanQCA};
they are a natural mathematical setting in which to encode the continuous-time and space, sometimes complex, behaviour of elementary physical elements (such as wave equations of particles, possibly interacting, under some field, for example) into the more discrete-data discrete-time framework of quantum computation. This issue of encoding is non-trivial, but has largely been solved for the one-dimensional case, by using QCA as a mathematical framework in which to model the system of interest. For example, several works \cite{Bialynicki-Birula,BoghosianTaylor2,BoghosianTaylor1,Eakins,LoveBoghosian,McGuigan,MeyerQLGI,MeyerQLGII,PerezCheung,Svozil,Vlasov} simulate quantum field theoretical equations in the continuous limit of a QCA dynamics. These results do not extend to more-than-one dimensions due to the problem of isotropy. Whilst in one-dimension the division of space into a line of points is an acceptable model, because it does not privilege the left over the right direction, in two-dimensions the grid seems to inevitably favour the cardinal directions (N, E, S, W) over the diagonal ones. That this construction is similar to the Billiard Ball Model CA  suggests a Quantum Billiard-Ball Model could be defined, moving away from the underlying grid, which would remedy this problem. This is planned this for future work.

After this work was completed, a two-dimensional, continuous-time, quantum version of the Game of Life was reported (see \href{http://arxiv.org/abs/1010.4666}{arXiv:1010.4666}).

\section*{Acknowledgements}
The authors would like to thank  J\'er\^ome Durand-Lose, Jarkko Kari, Simon Martiel, Kenichi Morita, Guillaume Theyssier and Philippe Jorrand.

\bibliography{../../../Bibliography/biblio}
\bibliographystyle{plain}

\end{document}